\pdfoutput=1
\documentclass[12pt]{article}
\usepackage[utf8]{inputenc}
\usepackage[british]{babel}
\usepackage{cmap}
\usepackage{lmodern}

\usepackage{amssymb, amsmath, amsthm}
\usepackage[a4paper,top=25mm,bottom=25mm,left=25mm,right=25mm]{geometry}
\usepackage{ragged2e}

\usepackage{authblk} 
\usepackage{pifont}
\usepackage{graphicx}
\usepackage[dvipsnames,svgnames,table]{xcolor}
\usepackage[figuresright]{rotating}
\usepackage{xtab} 
\usepackage{longtable} 
\usepackage{multirow}
\usepackage{footnote}
\usepackage[stable]{footmisc}
\usepackage{chngpage} 
\usepackage{pdflscape} 
\usepackage{tocbibind} 

\usepackage{pgfplots}
\pgfplotsset{compat=1.18}
\pgfplotsset{every tick label/.append style={font=\footnotesize}}
\usepackage{setspace}
\usepackage{pgfplotstable}
\usepackage{adjustbox}

\makesavenoteenv{tabular}
\usepackage{tabularx}
\usepackage{multirow}

\usepackage{booktabs}
\usepackage{threeparttable} 
\usepackage[referable]{threeparttablex} 
\newcolumntype{R}{>{\raggedleft\arraybackslash}X}
\newcolumntype{L}{>{\raggedright\arraybackslash}X}
\newcolumntype{C}{>{\centering\arraybackslash}X}
\newcolumntype{A}{>{\columncolor{gray!25}}C}
\newcolumntype{a}{>{\columncolor{gray!25}}c}

\newlength{\tablen}

\usepackage{dcolumn} 
\newcolumntype{.}{D{.}{.}{-1}}

\usepackage{tikz}
\usetikzlibrary{arrows, calc, matrix, patterns, positioning, trees}
\usepackage[semicolon]{natbib}
\usepackage[hyphens]{url}
\usepackage{hyperref} 
\hypersetup{
  colorlinks   = true,    		
  urlcolor     = blue,    		
  linkcolor    = blue,			
  citecolor    = ForestGreen,		  	
}
\usepackage{microtype}
\usepackage[justification=centering]{caption} 

\usepackage[labelformat=simple]{subcaption}

\DeclareCaptionLabelFormat{parenthesis}{(#2)}
\makeatletter
\renewcommand\p@subfigure{\arabic{figure}.}
\makeatother

\DeclareCaptionLabelFormat{parenthesis}{(#2)}
\makeatletter
\renewcommand\p@subtable{\arabic{table}.}
\makeatother

\usepackage{alphalph}

\def\addlegendimage{\csname pgfplots@addlegendimage\endcsname}

\usepackage{enumitem}

\setlist[itemize]{leftmargin=2.5\parindent}
\setlist[enumerate]{leftmargin=2.5\parindent}

\theoremstyle{plain}

\theoremstyle{definition}

\newtheorem{definition}{Definition}[section]

\theoremstyle{remark}

\def\keywords{\vspace{.5em} 
{\noindent \textit{Keywords}: }}

\def\AMS{\vspace{.5em} 
{\noindent \textbf{\emph{MSC} class}: }}

\def\JEL{\vspace{.5em} 
{\noindent \textbf{\emph{JEL} classification number}: }}

\title{The impact of the European Union’s enlargement with the Western Balkans and the Association Trio on the power of member states in the Council}
\author{
T\'imea Kov\'acs\thanks{
~E-mail: \emph{kvcs.timi@gmail.com} \newline
Budapest Metropolitan University, Budapest, Hungary} 
$\qquad \qquad$
\href{https://sites.google.com/view/doragretapetroczy}{D\'ora Gr\'eta Petr\'oczy}\thanks{~Corresponding author\newline 
E-mail: \emph{apetroczy@metropolitan.hu} \newline MNB Institute, Budapest Metropolitan University, Budapest, Hungary}
$\qquad \qquad$
G\'abor P\'asztor\thanks{
~E-mail: \emph{gpasztorhun@gmail.com} \newline
Prime Minister's Office, Budapest, Hungary} 
$\qquad \qquad$

}
\date{\today}

\begin{document}

\maketitle
\begin{abstract}
\noindent
As of 2022, the European Union has taken several steps regarding enlargement. We focus on the accession of countries with which the Union is actively negotiating membership. This is examined under two enlargement scenarios: first, the enlargement along the lines of the Western Balkan countries, and second, the accession of a trio (Ukraine, Moldova, and Georgia) to the already enlarged Union. We determine the a priori power of the member states based on the Banzhaf and Shapley--Shubik indices. Various coalitions are also assumed to assess the power and influence of member states, considering both pre- and post-enlargement scenarios. We found a rare case when the two indices give different rankings. In the case of the Western Balkan countries’ accession, the smaller population member states gain power, presenting an example of the new member paradox. While in a Union of 36 members, every member state loses some of their current power.  However, some coalitions are better off with the EU36 enlargement than a EU33 one.
\end{abstract}

\AMS{91A80, 91B12}

\JEL{C71, D72}

\section{Introduction}
The issue of enlargement has long been on the European Union's (EU) agenda. Currently, nine candidate countries are awaiting accession, ranging from Turkey, the longest-standing candidate, to Georgia, the most recent. In recent years, however, enlargement has gained new momentum due to the Russia–Ukraine conflict, as Ukraine and Moldova were granted official candidate status in 2022. These developments have caused dissatisfaction in the Western Balkans, where countries have been preparing for accession in close cooperation with the EU for years, while Ukraine has applied for accelerated accession within just a year and a half. This raises the question: will there be enlargement in the coming years, and if so, what impact will it have on the Union?

While enlargement would affect the EU’s daily functioning in many ways,  we examine how it would influence the balance of power and the relative strength of member states in the Council of the European Union.  To answer this question raises further issues. First: How much power do member states currently hold, and which countries are the most influential? To approach reality more closely, we also examine the power of several politically relevant coalitions: the V4, the Franco--German alliance, the Weimar Triangle, the founding members, the Nordic countries, and the 2004 accession group. Ultimately, we seek to understand how enlargement would affect these coalitions and who the winners and losers of an enlarged Union would be.

To determine power dynamics and the influence of countries, our analysis is based on the voting mechanism of the Council of the European Union. The Council and the Parliament are co-legislators; however, the Council is where national ministers represent their member states. Therefore, a member state's power is primarily determined by its ability to influence or block decisions. 

According to qualified majority voting (QMV), a proposal is adopted if:
\begin{itemize}
\item at least 55\% of the member states (currently 15) vote in favour, 
\item and these states represent at least 65\% of the EU population.
\end{itemize}
Furthermore, a blocking minority must include at least four Council members; failing this, the qualified majority shall be deemed to have been attained.

We consider two enlargement scenarios. First, we examine the accession of the Western Balkan countries (EU33), followed by the accession of the Association Trio, the most recently granted candidate countries (EU36). The latter group includes Georgia, Moldova, and Ukraine. Turkey is excluded from the study, as accession negotiations have stalled since 2018.

We apply a game-theoretical methodology. This approach is widely used in the literature; a priori voting power is usually measured by the Banzhaf and Shapley--Shubik indices. These indices quantify the influence of countries in decision-making by assigning a value to each country based on its ability to affect Council decisions. They illustrate the statistical aspect of voting possibilities, but do not encompass the political ties between member states, instead treating them as homogeneous.

Unsurprisingly, the largest countries lose the most power in both the EU33 and EU36 scenarios. However, in the EU33 scenario, the phenomenon known as the \emph{new member paradox} appears, meaning that while the power of the largest countries decreases with a new member, the power of the countries with the smallest populations increase \citep{BramsAffuso1976}. Thus, the accession of the Western Balkans would benefit the smallest countries while negatively impacting the largest countries.

In contrast, the EU36 scenario is disadvantageous for all current members; however, some coalitions lose less power in the case of EU36 than in EU33 compared to the present situation. Ukraine would gain a strong position, becoming the country with the fifth-largest power. Such a significant power of a new member state could result in new dynamics in the balance of power within the EU.

The structure of our study is as follows. Section~\ref{Sec2} provides a concise overview of related papers.
Section~\ref{Sec3} introduces the basic concepts of our approach, which are used to carry out the numerical analysis in Section~\ref{Sec4}. Finally, Section~\ref{Sec5} ends with concluding remarks.

\section{Related literature} \label{Sec2}

The Council of the European Union, as a well-known case of weighted voting, is the subject of several academic studies. We limit our discussion to a few salient examples that are closely tied to our research. The connecting literature centres on three main topics: the examination of voting weights, the exit of a member state, and enlargement.

\subsection{Voting weights}

In their seminal works, \citet{BramsAffuso1976, BramsAffuso1985} investigate the paradoxes of voting power in weighted voting systems, particularly within the context of institutional expansions. 
\citet{BramsAffuso1976} probably provide the first academic demonstration that Luxembourg was a null player in the Council of Ministers of the European Economic Community, the predecessor of the European Union. Between 1958 and 1973, the weights of France, Germany, and Italy were 4, Belgium and the Netherlands had a weight of 2, while Luxembourg had a weight of 1. The qualified majority required 12 out of the 17 votes; hence, Luxembourg had absolutely no power.

 \citet{BramsAffuso1976} also introduce the `paradox of new members', demonstrating through theoretical models and empirical cases—such as the U.S. Electoral College and the European Community (EC) Council of Ministers—that the addition of new members to a voting body can counterintuitively increase the voting power of some existing members, even as their share of total votes declines. They analyse this phenomenon using several indices of voting power, notably the Banzhaf and Shapley--Shubik, and develop efficient computational methods via generating functions to assess its frequency and conditions. The follow-up study \citet{BramsAffuso1985} extends this analysis to further EC expansions, revealing that Luxembourg, despite its minimal vote share, gained voting power with each enlargement, and even achieved parity with larger states, such as Denmark and Ireland, under certain decision rules. These findings underscore the difference between voting weight and actual influence, highlighting the need for formal voting-power analysis in institutional design to prevent unintended and politically significant anomalies.

\citet{AlonsoMeijideBilbaoCasasMendezFernandez2009} introduce computational methods based on generating functions to calculate coalitional power indices in weighted multiple majority games, particularly when players are organised into predefined coalition structures. Their findings highlight how coalition structures and voting rules affect the distribution of power among EU member states.

\citet{NapelWidgren2011} compare the results from traditional voting power indices to a strategic voting power assessment. The differences between the relative power indicated by the strategic approach and standard indices turn out to be quite close if the European Commission’s aggregate policy ideals are assumed to be drawn independently from the same distribution as those of the Council members. In other words, traditional indices provide a good approximation of power even if strategic considerations are also considered.

\citet{Koczy2012} looks at the immediate impact of the Lisbon Treaty reform in 2014, as well as at the long-term effects of demographic trends in the countries. The reform is demonstrated to hurt medium-sized countries, especially states with declining populations.

\citet{LeBretonMonteroZaporozhets2012}  critically examine how voting power has been distributed in the EU Council of Ministers from 1958 onward. The authors argue that traditional power indices like the Banzhaf and Shapley–Shubik indices are inadequate for distributive settings and propose the nucleolus as a more robust and fair measure of power. They compute the nucleolus for various historical EU voting rules and compare it to population shares to assess fairness. The analysis reveals that voting power often diverges from population-based fairness, especially disadvantaging larger countries like Germany. They also propose a methodology to design optimal voting rules that minimise inequality among EU citizens, showing that fairer allocations were possible in earlier EU configurations.

\citet{PetroczyCsato2023} focus on how changes to the population and member state thresholds affect power dynamics. They conduct a detailed sensitivity analysis across a range of plausible threshold values and find that voting power is highly sensitive to these parameters. By appropriately modifying quotas, decision-making power can be increased, while the voting power remains essentially unchanged.

\subsection{Exit of member states from the European Union}

The withdrawal of the United Kingdom (Brexit) was the first instance of a member state leaving the Union in 2016.
Several studies have shown independently that it has mainly benefited large countries \citep{Gabor2020, Grech2021, Gollner2018, Kirsch2016, Koczy2021, MercikRamsey2017}. On the other hand, a further exit would harm the large and benefit the small countries due to the unchanged states criterion at 15 \citep{PetroczyRogersKoczy2022}. \citet{KirschSlomczynskiStolickiZyczkowski2018} use the normal approximation of the Banzhaf index in double-majority games to uncover that such non-monotonicity is, in most cases, inherent in a double-majority system. However, it is strongly exacerbated by the peculiarities of the population vector in the European Union.

\citet{Szczypinska2018} classifies EU member states into coalitions based on several criteria. The analysis is distinguished between countries that use the euro and those that do not; countries experiencing macroeconomic imbalances, including excessive ones, versus those without such issues; net contributors to the EU budget versus net beneficiaries; and countries that are proponents, moderate supporters, or sceptics of EU regulations.  Following Brexit, non-eurozone countries would face increased difficulty in forming a blocking minority. Conversely, the power position of net contributor states remains largely unaffected by the UK's departure—they continue to hold a strong position within the EU decision-making framework.

\subsection{EU enlargement}

\citet{HerneNurmi1993} examine the accession of Finland, Sweden, Norway, and Austria to the European Economic Community under the rules in force in 1993. Based on these rules, they assign appropriate weights to each of the four countries according to their populations and find that Austria's accession would benefit Luxembourg and Sweden the most. Furthermore, if 4 votes are assigned to Sweden instead of 5, it would be advantageous for Sweden, Denmark, Finland, and even Ireland. Additionally, \citet{HerneNurmi1993} use both power indices in their calculations and highlight the differences between them. However, since these differences are not dramatic, they recommend applying both indices.

\citet{Widgren1994} calculates the effect of two possible enlargements, the potential joining of Sweden, Austria, Finland, and Norway (EC16), as well as Switzerland, Iceland, and Liechtenstein (EC19). Although these enlargements have never occurred, the study considers not only the Banzhaf and Shapley--Shubik indices but also possible \emph{a priori} coalitions (France--German axis, Benelux countries, Mediterranean countries, Nordic countries). \citet{FelsenthalMachover1997} analyse the impact of four historical enlargements (1973, 1981, 1985, 1995), and determine a responsiveness index of the voting rule, which is a measure of decision ability.

In the history of the EU, the Treaty of Nice (2001) was the first that take the populations of the member states into account. A proposal could have been adopted if the supporting countries had at least 74\% of the weights and 62\% of the population.
According to \citet{FelsenthalMachover2001}, these quotas were fair, as each voter had approximately the same power in the Council; however, they were excessively high and essentially paralysed decision-making.
\citet{Leech2002} essentially reproduces the work of \citet{FelsenthalMachover2001}, and proposes a fair quota system before the 2004 EU enlargement, in which all citizens have roughly equal power on decisions, regardless of the number of member states. The suggested algorithm adjusts the weights until the relative voting power and the population share are sufficiently close. \citet{SlomczynskiZyczkowski2006} investigate a similar system based on the law of Penrose, where the weight of each country is proportional to the square root of its population.

\citet{AleskerovAvciIakoubaTurem2002} focus on the implications of future enlargements by using the Banzhaf and Shapley--Shubik indices to evaluate past and emerging power distributions in the Council of the European Union under the Treaty of Nice. The power of the larger (smaller) countries is found to be lower (higher) with the Banzhaf index compared to the Shapley--Shubik index.

\citet{Mylona2007} examines the countries of the Western Balkans, meaning Albania, Bosnia and Herzegovina, Croatia, Montenegro, North Macedonia, and  Serbia. At that time, the Nice rules were in force, but the 2004 EU constitution already proposed a reform for the Council's voting process. The calculations are based on the Shapley--Shubik index in both scenarios. It is concluded that under the Nice Treaty, the EU's efficiency would significantly decrease, and enlargement is not recommended until a new decision-making reform is in place. According to the new reform's calculations, enlargement would not weaken efficiency, but smaller countries would lose some of their power.

\citet{Kirsch2022} investigates the impact of the accession of Montenegro, Turkey, and Ukraine on the current 27 member states. In this case, the 55\% states threshold will grow from 15 to 16. Consequently, small countries' power would increase, and large countries' power would decrease. Due to its substantial size, the accession of Turkey would tip the balance of power: France, Poland, Germany, Italy, and Spain would lose nearly 20\% of their voting power.

In the working paper, \citet{Ehin2025} analyse how potential EU enlargements—specifically the accession of the Association Trio (Ukraine, Moldova, Georgia), the Western Balkans, or both would shift voting weights in the Council of the EU. Using ratio calculations for member states and populations, it is found that enlargement would increase the voting weights of smaller, newer, Eastern, and Southern states, while reducing the relative power of older, larger, and Western states. Despite this, large and founding members would still be able to block decisions. Compared to our study, this working paper examines only the ratio of voting weight, not power indices.

\section{Methodology} \label{Sec3}
Voting situations are usually modeled by a cooperative game with transferable utility where the voters are the players, and the value of any coalition is maximal if it can accept a proposal and minimal otherwise.

Let $N$ denote the set of players and $S \subseteq N$ be a coalition. The cardinality of a set is denoted by the corresponding small letter, namely, the number of players in coalition $S$ is $\lvert S \rvert = s$, and the number of players is $\lvert N \rvert = n$. The value of any coalition is given by the characteristic function $v: 2^N \to \mathbb{R}$.

\begin{definition}
\emph{Simple voting game}:
A game $(N,v)$ is called a simple voting game if
\[
v(S) \in \left\{ 0,1 \right\}  \text{ for all $S \subseteq N$}.
\]
\end{definition}

Let vector $\mathbf{w} \in \mathbb{R}^n$ denote the weights of the players and $q \in \left[ 0,\, 1 \right]$ denote the decision threshold.

\begin{definition}
\emph{Weighted voting game}:
A game $(N,v,\mathbf{w},q)$ is called a weighted voting game if for any coalition $S \subseteq N$:
\[
v(S)= \left\{ 
\begin{array}{cl}
1 & \text{if }   \sum_{j \in S} w_j \geq q \\ 
0 & \text{otherwise}.
\end{array}
\right.
\]
\end{definition}

The qualified majority voting system of the EU Council can be formalised as a combination of three weighted voting games \citep[p.~1248]{KurzNapel2016}. In particular, the associated characteristic function is $v = \left( v^{(1)} \land v^{(2)} \right) \lor v^{(3)}$, where
\begin{itemize}
\item
$v^{(1)} = \left[ \mathbf{w}^{(1)}, q^{(1)} \right]$ such that $w_i^{(1)}$ is the population of country $i$ for all $1 \leq i \leq 27$ and $q^{(1)}$ is 65\% of the total population in the EU (population criterion);
\item
$v^{(2)} = \left[ \mathbf{w}^{(2)}, q^{(2)} \right]$ such that $w_i^{(2)} = 1$ for all $1 \leq i \leq 27$ and $q^{(2)} = 15$ (states criterion);
\item
$v^{(3)} = \left[ \mathbf{w}^{(3)}, q^{(3)} \right]$ such that $w_i^{(3)} = 1$ for all $1 \leq i \leq 27$ and $q^{(3)} = 24$ (blocking minority rule).
\end{itemize}

However, it is worth mentioning that the blocking minority rule has almost no influence on the value of the power indices. \citet[Appendix E]{PetroczyRogersKoczy2022} find only 19 coalitions of 24 states, which do not represent 65\% of the total population of the EU.

A player is called \textit{critical} if it can turn a winning coalition into a losing one. The  Banzhaf index, which is the normalised Banzhaf value \citep{Banzhaf1964, Coleman1971, Penrose1946}, shows the probability that a player influences a decision by taking into account the ratio of coalitions where the player is critical. As we have already seen, the Banzhaf index is one of the most common measures of a priori voting power 

\begin{definition}
Let $(N,v)$ be a simple voting game.
The \emph{Banzhaf value} of player $i \in N$ is
\[
\sum_{S  \subseteq N \setminus  \{i\}}  \frac{1}{2^{n-1} }\left(v\big(S\cup\left\{i\right\}\big)-v\big(S\big)\right)= \frac{\eta_i (N, v)}{2^{n-1}},
\]
where $\eta_i(v)$ is the Banzhaf score of player $i$, the number of coalitions where $i$ is critical.
\end{definition}

Usually, its normalised version is reported as the measure of voting power.

\begin{definition}
Let $(N,v)$ be a simple voting game.
The \emph{Banzhaf index} of player $i \in N$ is its normalised Banzhaf score:
\[
\beta_i(N,v) =\frac{\eta_i (N,v)}{\sum_{j\in N} \eta_j (N,v)}.
\]
\end{definition}

In the literature, the Banzhaf index is sometimes called Normalised Banzhaf Index (\textit{NBI}) to emphasise that it is a normalised value. 

Consider a random order of the players. The Shapley--Shubik index of a player is its marginal contribution to the coalition formed by the preceding players, averaged over the set of all the possible orders of the players \citep{Shapley1953,ShapleyShubik1954}.

\begin{definition}
\emph{Shapley--Shubik index}:
Let $(N,v)$ be a simple voting game.
The Shapley--Shubik index of player $i$ is
\[
\varphi_i(N,v) = \sum_{S  \subseteq N \setminus  \{i\}}  \frac{s!(n-s-1)!}{n!} \left[ v \big( S \cup \left\{ i \right\} \big) -v \left( S \right) \right].
\]
\end{definition}


\citet{FelsenthalMachover1998} and \citet{FelsenthalMachover2004} distinguish between two intuitive notions of what voting power means, leading to two approaches to measure it. The first concept, I-power, focuses on a voter's potential influence over the outcome of decisions by a voting body. The second concept, P-power, focuses on voters' payoff, their expected share of a fixed winning `prize’. Banzhaf’s approach corresponds to the I-power, while Shapley–Shubik’s is a P-power notion. Since both indices have their own justification, we performed our calculations using both.

In many cases, the Shapley–Shubik and Banzhaf indices have fairly similar values, but in general they can behave quite differently \citep[pp. 277-278.]{FelsenthalMachover1998}. However, \citet{FreixasMarciniakPons2012} prove ordinal equivalence of these two indices in the class of semicomplete games.

To calculate these indices, we use the \href{https://pypi.org/project/power-bdd/}{power-bdd Python package}, which implements the algorithm described in \citet{Bolus2011}. This calculates the power indices by writing the game in binary decision diagram form. \citet{Wilms2020} verifies that this method works much faster in certain situations than other dynamic programming algorithms. For the sake of simplicity, we omit the blocking minority rule from the calculation; according to \citet{PetroczyRogersKoczy2022} and \citet{Kirsch2022}, this essentially does not influence the results.

\subsection{The choice of candidate countries}
We include all current candidate states (and Kosovo, which has potential candidate status) with whom the EU actively cooperates to promote accession. The research does not cover Turkey, which became an official candidate in 1999, as its accession negotiations stalled in 2018 \citep{EuropeanCouncil2024}. Furthermore, Kosovo's situation is unclear, as five EU member states do not recognise its independence – Cyprus, Greece, Romania, Slovakia, and  Spain  \citep{EuropeanParliament2024}, thus it can only be a potential candidate. However, the Union treats Kosovo within the same framework as other Western Balkan candidate countries and actively negotiates to promote Kosovo's accession to the European Union.

The candidate states covered in the paper are divided into two groups. The first group is the Western Balkan states, which are Albania, Bosnia and Herzegovina, Kosovo, Montenegro,  North Macedonia, and Serbia. The second group is the newly candidate states, which are Georgia, Moldova, and Ukraine.

There is an active debate in the Union about the merits of gradual or unified enlargement. We decided to examine the unified accession of the two defined groups, because the 2004 enlargement set a precedent for this. At that time, Hungary and Poland were far ahead in accession compared to the other eight candidates, but the decision was made to carry out the enlargement simultaneously. Based on this, we treat the Western Balkan countries and the new candidate states as separate units in the paper.

First, we examine the accession of the Western Balkan countries, as the most likely candidates for the next accession. In June 2022, EU leaders expressed their full and unequivocal commitment to the EU membership perspective of the Western Balkans and urged the acceleration of the accession process \citep{EuropeanParliament2024}. In addition to this commitment, they also assist in reforms and financial support within the framework of the Stabilisation and Association Process (SAP) and the new Growth Plan. Regular summits are held to review the processes.

Next, we examine Georgia, Moldova, and Ukraine, which gained candidate status under the pressure of the Russian invasion of Ukraine in recent years. The EU emphasised merit-based accession for the Western Balkan countries. This credibility was disrupted by Moldova and Ukraine receiving candidate status within one year of submitting their accession application, which was unprecedented, and it was clear that the decision to grant candidate status was politically motivated. It is important to contrast them with the Western Balkan countries, which have been striving to meet the conditions for decades, but lacked the political pressure that could have brought about accession.

We do not examine the two enlargement scenarios separately, but in a given order. That is, Georgia, Moldova, and Ukraine would not join the EU27, but the already expanded EU33 with the Western Balkans. The reason for this is the advancement of the Western Balkan countries in accession negotiations.

\subsection{Coalitions}

During our study, we establish some coalitions to provide a more comprehensive picture of the power dynamics within the EU and how they would evolve as a result of enlargement. Naturally, the member state weight of a coalition equals the number of countries in the coalition; for all other countries, the weight remains 1. Similarly, the population quota of a coalition is the sum of the populations of its member countries. 

The first coalition is the Franco--German cooperation, which plays a key role in the formation of the Union and the deepening of European integration. These two countries are not only the largest in the Union in terms of population and economic performance, but also committed shapers of EU policies. Their cooperation is often referred to as the Franco--German axis \citep{KaedingSelck2005}.

Next, we analyse the Weimar Triangle, a cooperation between France, Germany, and Poland. Established in 1991, this forum was created to foster collaboration based on European values after the Cold War, support Poland’s integration into NATO and the EU, and deepen the Franco--German relationship \citep{Maurice2022}. In recent years, their cooperation has strengthened, particularly in matters concerning Ukraine, energy policy, and EU defence policy \citep{Bendiek2008, LangSchwarzer2011, Plociennik2020}. 

We then examine the founding members of the EU’s core: Belgium, France, Germany (then West Germany), Italy, Luxembourg, and the Netherlands. These six countries founded the European Coal and Steel Community in 1951 to place their coal and steel industries under joint control and prevent future armed conflict \citep{EuropeanUnion2024a}. Later, they also established the European Economic Community, the predecessor of the EU, making their power position within the Union crucial.

During the Cold War, Europe was divided into Western and Eastern blocs, which diverged ideologically and economically \citep{EuropeanUnion2024b}. After the Cold War, Eastern countries underwent gradual integration, culminating in their accession in 2004. Therefore, we examine the coalition of the 10 countries that joined in 2004, representing the largest single enlargement and the Eastern bloc within the EU. We also assessed their power relative to the founding members.

We then analysed the Visegrád Group (V4), which includes the Czech Republic, Hungary, Poland, and Slovakia. The V4 has seen success in areas such as EU neighbourhood policy, energy security, and transport coordination. However, internal political disagreements often hinder cooperation \citep{Walsch2014}. In recent years, they have acted jointly on migration issues, but diverging views on the Russian invasion of Ukraine have strained the alliance.

Finally, we examine the regional coalition of the Nordic countries: Denmark, Estonia, Finland, Latvia, Lithuania, and Sweden. Their cooperation is most visible in fisheries, energy policy, agriculture, and EU foreign relations. The Nordic Council provides a platform for this collaboration, with Denmark, Finland, Iceland, Norway, and Sweden as members, and Estonia, Latvia, and Lithuania as observers \citep{NordicCooperation}.

\section{Results} \label{Sec4}

This section presents the development of the analysis and the results. First, we examine the current balance of power within the European Union. In the Union, countries are most effective in achieving their goals collectively; therefore, we compute the power indices of various politically relevant coalitions within the current EU27. Then we compare these results with two enlargement scenarios: first with the Western Balkan countries (EU33), and then with the new candidate countries joining the EU33 (EU36), namely, Georgia, Moldova, and Ukraine.


\subsection{EU27 power relations}

First, we examine the state of power relations within the European Union in 2024 through the Council’s voting mechanism. To calculate the power indices of the 27 member states, we use the normalised Banzhaf index and the Shapley--Shubik index. The population data for 2023 comes from the open-access Eurostat database \citep{Eurostat2023}.

Both the Banzhaf and the Shapley--Shubik indices produce similar results for the EU27. The differences between them stem solely from methodological variations; however, they both yielded the same ranking of countries in terms of power. This ranking corresponds to the population data, as a country with a higher population has more power.

\begin{table}[t!]
\centering
\caption{Power indices of the most populous EU27 countries}
\begin{tabularx}{1.05\textwidth}{lCcc}
\toprule
Country & Population (\%) & Norm. Banzhaf index (\%) & Shapley--Shubik index (\%) \\
\bottomrule
\rowcolor{gray!25} Germany       & 18.81 & 12.21 & 18.14 \\
France        & 15.18 & 10.08 & 13.60 \\
\rowcolor{gray!25} Italy         & 13.12 & 8.77  & 11.48 \\
Spain         & 10.72 & 7.69  & 9.27  \\
\rowcolor{gray!25} Poland        & 8.20  & 6.21  & 6.64  \\
\toprule
\end{tabularx}
\label{tab1}
\end{table}

As shown in Table~\ref{tab1}, Germany and France hold the most significant influence according to both power indices. In the case of the Banzhaf index, Germany scores 12.21\% and France 10.08\%. Meanwhile, the Shapley--Shubik index assigns 18.14\% to Germany and 13.6\% to France. These power values are relatively high, just like the population figures of the two countries, which together account for nearly 34\% of the EU’s total population. The balance of power between these two countries within the EU is often referred to as the Franco--German axis. The presence of this axis is evidenced by the fact that most proposals submitted to the Commission originate from French or German initiatives. Furthermore, it is worth highlighting that without the support of these two member states, it is hardly possible to adopt new rules or pass proposals within the EU.


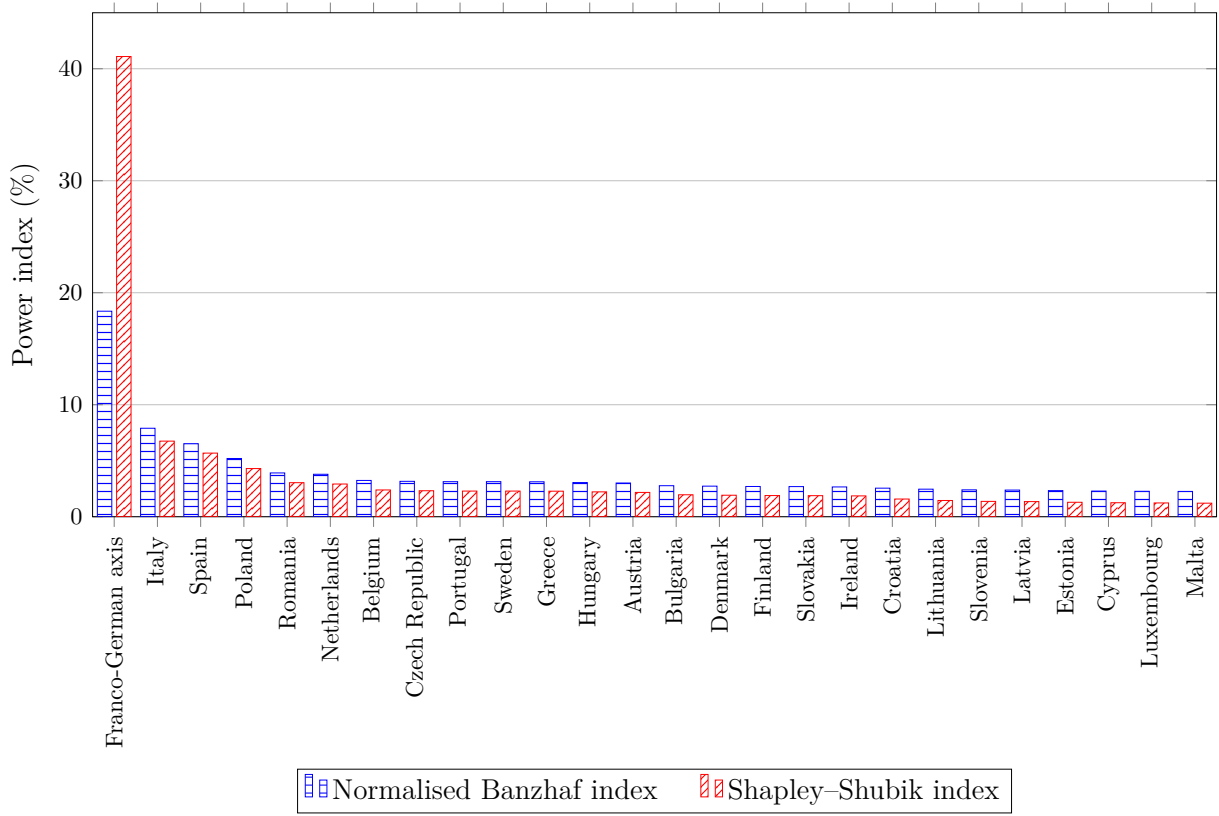
\begin{figure}[!t]
\centering
\begin{adjustbox}{width=\textwidth}
\begin{tikzpicture}
\begin{axis}[
    ybar,
    bar width=6pt,
    width=18cm,
    height=9cm,
    enlarge x limits=0.02,
    legend style={at={(0.5,-0.5)}, anchor=north, legend columns=-1},
    ylabel={Power index (\%)},
    symbolic x coords={Franco-German axis,Italy,Spain,Poland,Romania,Netherlands,Belgium,Czech Republic,Portugal,Sweden,Greece,Hungary,Austria,Bulgaria,Denmark,Finland,Slovakia,Ireland,Croatia,Lithuania,Slovenia,Latvia,Estonia,Cyprus,Luxembourg,Malta},
    xtick=data,
    x tick label style={rotate=90, anchor=east},
    ymin=0,
    ymax=45,
    ymajorgrids = true,
    legend cell align={left}
]
\addplot+[style={blue, fill=blue!30, pattern=horizontal lines,  pattern color=blue}] coordinates {
(Franco-German axis,18.35)
(Italy,7.9)
(Spain,6.52)
(Poland,5.2)
(Romania,3.91)
(Netherlands,3.79)
(Belgium,3.24)
(Czech Republic,3.16)
(Portugal,3.13)
(Sweden,3.13)
(Greece,3.12)
(Hungary,3.05)
(Austria,3.01)
(Bulgaria,2.77)
(Denmark,2.73)
(Finland,2.7)
(Slovakia,2.69)
(Ireland,2.66)
(Croatia,2.55)
(Lithuania,2.46)
(Slovenia,2.4)
(Latvia,2.38)
(Estonia,2.33)
(Cyprus,2.29)
(Luxembourg,2.27)
(Malta,2.26)
};
\addplot+[style={red, pattern=north east lines, pattern color=red}] coordinates {
(Franco-German axis,41.09)
(Italy,6.75)
(Spain,5.68)
(Poland,4.3)
(Romania,3.04)
(Netherlands,2.92)
(Belgium,2.39)
(Czech Republic,2.32)
(Portugal,2.29)
(Sweden,2.29)
(Greece,2.28)
(Hungary,2.22)
(Austria,2.17)
(Bulgaria,1.96)
(Denmark,1.92)
(Finland,1.89)
(Slovakia,1.88)
(Ireland,1.86)
(Croatia,1.58)
(Lithuania,1.45)
(Slovenia,1.37)
(Latvia,1.35)
(Estonia,1.29)
(Cyprus,1.25)
(Luxembourg,1.23)
(Malta,1.21)
};
\legend{Normalised Banzhaf index \quad\quad, Shapley--Shubik index}
\end{axis}
\end{tikzpicture}
\end{adjustbox}
\caption{Power indices with French and German cooperation}
\label{Figure1}
\end{figure}



\begin{figure}[!t]
\centering
\begin{adjustbox}{width=\textwidth}
\begin{tikzpicture}
\begin{axis}[
    ybar,
    bar width=6pt,
    width=18cm,
    height=9cm,
    enlarge x limits=0.02,
    legend style={at={(0.5,-0.5)}, anchor=north, legend columns=-1},
    ylabel={Power index (\%)},
    symbolic x coords={Weimar,Italy,Spain,Romania,Netherlands,Belgium,Czech Republic,Portugal,Sweden,Greece,Hungary,Austria,Bulgaria,Denmark,Finland,Slovakia,Ireland,Croatia,Lithuania,Slovenia,Latvia,Estonia,Cyprus,Luxembourg,Malta},
    xtick=data,
    x tick label style={rotate=90, anchor=east},
    ymin=0,
    ymax=55,
    ymajorgrids = true,
    legend cell align={left}
]
\addplot+[style={blue,fill=blue!30, pattern=horizontal lines,  pattern color=blue}] coordinates {
(Weimar,21.74)
(Italy,7.04)
(Spain,6.88)
(Romania,3.52)
(Netherlands,3.47)
(Belgium,3.19)
(Czech Republic,3.14)
(Greece,3.12)
(Portugal,3.12)
(Sweden,3.12)
(Hungary,3.07)
(Austria,3.05)
(Bulgaria,2.9)
(Denmark,2.88)
(Finland,2.86)
(Slovakia,2.85)
(Ireland,2.84)
(Croatia,2.76)
(Lithuania,2.71)
(Slovenia,2.67)
(Latvia,2.66)
(Estonia,2.63)
(Cyprus,2.61)
(Luxembourg,2.59)
(Malta,2.59)

};
\addplot+[style={red,fill=red!30, pattern=north east lines, pattern color=red}] coordinates {
(Weimar,49.6)
(Italy,5.19)
(Spain,5.1)
(Romania,2.48)
(Netherlands,2.41)
(Belgium,2.08)
(Czech Republic,2.03)
(Greece,2.01)
(Portugal,2.01)
(Sweden,2.01)
(Hungary,1.96)
(Austria,1.94)
(Bulgaria,1.79)
(Denmark,1.77)
(Finland,1.75)
(Slovakia,1.74)
(Ireland,1.73)
(Croatia,1.66)
(Lithuania,1.61)
(Slovenia,1.57)
(Latvia,1.55)
(Estonia,1.53)
(Cyprus,1.5)
(Luxembourg,1.49)
(Malta,1.48)

};
\legend{Normalised Banzhaf index \quad\quad, Shapley--Shubik index}
\end{axis}
\end{tikzpicture}
\end{adjustbox}
\caption{Power indices with the Weimar cooperation}
\label{Figure2}
\end{figure}
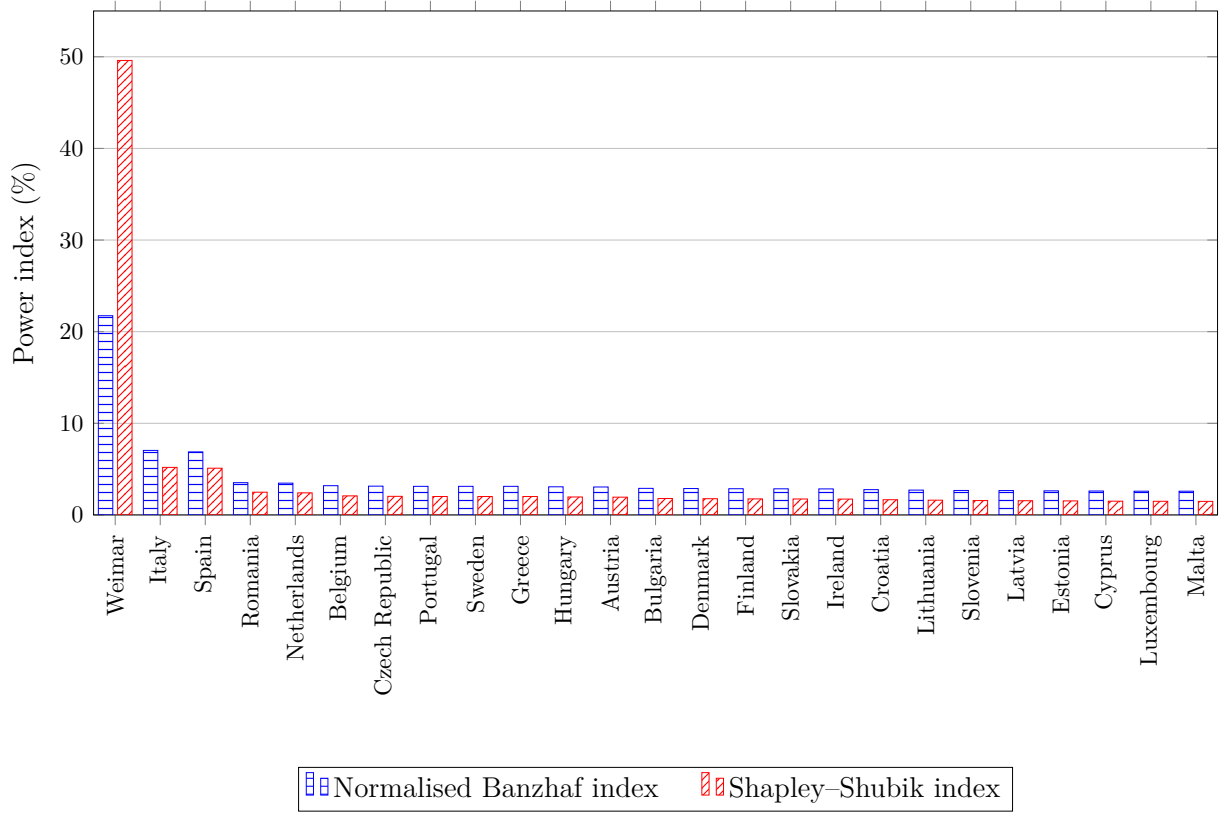


\begin{figure}[!t]
\centering
\begin{adjustbox}{width=\textwidth}
\begin{tikzpicture}
\begin{axis}[
    ybar,
    bar width=6pt,
    width=18cm,
    height=9cm,
    enlarge x limits=0.02,
    legend style={at={(0.5,-0.5)}, anchor=north, legend columns=-1},
    ylabel={Power index (\%)},
    symbolic x coords={Founder countries, Spain,Poland,Romania,Czech Republic,Greece,Portugal,Sweden,Hungary,Austria,Bulgaria,Denmark,Finland,Slovakia,Ireland,Croatia,Lithuania,Slovenia,Latvia,Estonia,Cyprus, Malta},
    xtick=data,
    x tick label style={rotate=90, anchor=east},
    ymin=0,
    ymax=60,
    ymajorgrids = true,
    legend cell align={left}
]
\addplot+[style={blue,fill=blue!30, pattern=horizontal lines,  pattern color=blue}] coordinates {
(Founder countries,37.99)
(Spain,3.58)
(Poland,3.58)
(Romania,3.52)
(Czech Republic,3.14)
(Greece,3.12)
(Portugal,3.12)
(Sweden,3.12)
(Hungary,3.07)
(Austria,3.04)
(Bulgaria,2.89)
(Denmark,2.86)
(Finland,2.84)
(Slovakia,2.83)
(Ireland,2.82)
(Croatia,2.75)
(Lithuania,2.68)
(Slovenia,2.65)
(Latvia,2.63)
(Estonia,2.61)
(Cyprus,2.58)
(Malta,2.56)
};
\addplot+[style={red,fill=red!30, pattern=north east lines, pattern color=red}] coordinates {
(Founder countries,58.64)
(Spain,2.35)
(Poland,2.35)
(Romania,2.32)
(Czech Republic,2.08)
(Greece,2.07)
(Portugal,2.07)
(Sweden,2.07)
(Hungary,2.04)
(Austria,2.02)
(Bulgaria,1.93)
(Denmark,1.91)
(Finland,1.9)
(Slovakia,1.9)
(Ireland,1.89)
(Croatia,1.85)
(Lithuania,1.81)
(Slovenia,1.78)
(Latvia,1.78)
(Estonia,1.76)
(Cyprus,1.75)
(Malta,1.73)

};
\legend{Normalised Banzhaf index \quad\quad, Shapley--Shubik index}
\end{axis}
\end{tikzpicture}
\end{adjustbox}
\caption{Power indices with the cooperation of founding countries}
\label{Figure3}
\end{figure}
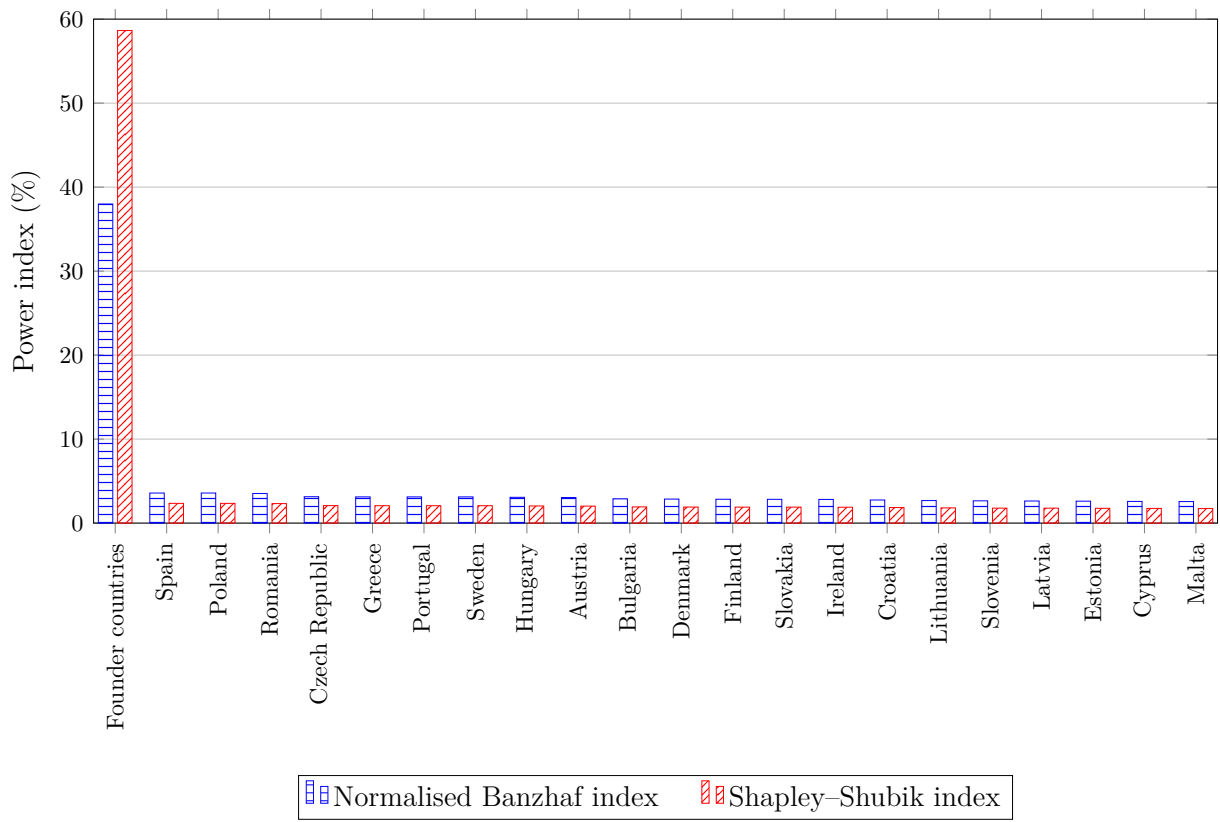

To better illustrate the influence of the two member states, we examine the scenario in which they make every decision together, that is, they form a coalition. This means they never vote differently. As shown in Figure~\ref{Figure1}, although the order of the indices remained unchanged, the Shapley--Shubik index values this setup at more than twice the level of the Banzhaf index. The difference is mostly balanced out among the smaller member states, as the Shapley--Shubik index assigns them less power. The discrepancy between the power indices reflects methodological differences. Nevertheless, both indices indicate that the two countries jointly wield significant influence in the EU, which is further amplified by their political and economic clout.

Among the EU member states, two southern countries—Italy and Spain—come closest to matching the power of Germany and France (see Table~\ref{tab1}). Their populations and power indices substantially exceed the EU average of 3.7\% for both the Banzhaf and Shapley--Shubik indices. Italy is one of the EU’s founding members, while Spain joined in 1986. Both countries hold considerable influence, yet neither aspires to lead or shape the EU according to their vision, unlike France or Germany.

The following most powerful country is Poland (see Table~\ref{tab1}). In terms of population, Poland plays a unique role in the EU, as it is closer in size to the larger member states, yet it is not considered part of the EU core. Poland is not a founding member and only joined in 2004. It has not adopted the euro as its official currency; thus, it is excluded from specific Eurozone decision-making processes, which results in a degree of marginalisation. Still, based on power indices, it holds the fifth-largest influence in the Council.

Germany and France have also recognised Poland’s unique position. To maintain good relations and support Poland’s western integration, they established the Weimar Triangle with Poland in 1991. This initiative promoted close cooperation among the three countries and created a platform for meetings between the heads of state and foreign ministers \citep{FranceDiplomacy2024}. Figure~\ref{Figure2} illustrates the outcome of these three countries cooperating in every decision. Unsurprisingly, they hold a substantial concentration of power compared to other countries. According to the Shapley--Shubik index, their combined power is nearly 50\%, while the Banzhaf index places it at over 20\%.

Three of the EU’s most powerful member states are founding members. Therefore, it is interesting to observe the significant power held by the original six founding countries: Belgium, France, Germany, Italy, Luxembourg, and the Netherlands. Figure~\ref{Figure3} shows the scenario in which all six founding members vote together, forming a coalition. It can be observed that their power index is nearly 40\% according to the Banzhaf index and 60\% according to the Shapley--Shubik index, while the power indices of all other member states range between 1\% and 4\%.

\begin{table}[t!]
\centering
\caption{Power indices of medium-population EU27 countries}
\rowcolors{0}{}{gray!25}
\begin{tabularx}{1.05\textwidth}{lCcc}\hiderowcolors
\toprule
Country & Population (\%) & Norm. Banzhaf index (\%) & Shapley--Shubik index (\%) \\
\bottomrule
\showrowcolors
Romania       & 4.25 & 3.94 & 3.81 \\
Netherlands   & 3.97 & 3.79 & 3.60 \\
Belgium       & 2.62 & 3.08 & 2.62 \\
Czech Republic& 2.41 & 2.97 & 2.47 \\
Sweden        & 2.35 & 2.93 & 2.43 \\
Portugal      & 2.33 & 2.93 & 2.42 \\
Greece        & 2.32 & 2.92 & 2.41 \\
Hungary       & 2.14 & 2.82 & 2.28 \\
Austria       & 2.03 & 2.76 & 2.21 \\
Bulgaria      & 1.44 & 2.45 & 1.78 \\
Denmark       & 1.32 & 2.39 & 1.70 \\
Finland       & 1.24 & 2.34 & 1.65 \\
Slovakia      & 1.21 & 2.33 & 1.63 \\
Ireland       & 1.16 & 2.30 & 1.59 \\
\toprule
\end{tabularx}
\label{Table2}
\end{table}

\begin{table}[t!]
\centering
\caption{Power indices of small population EU27 countries}
\rowcolors{0}{}{gray!25}
\begin{tabularx}{1.05\textwidth}{lCcc}\hiderowcolors
\toprule
Country & Population (\%) & Norm. Banzhaf index (\%) & Shapley--Shubik index (\%) \\
\bottomrule
\showrowcolors
Croatia       & 0.86 & 2.14 & 1.37 \\
Lithuania     & 0.64 & 2.02 & 1.21 \\
Slovenia      & 0.47 & 1.93 & 1.08 \\
Latvia        & 0.42 & 1.90 & 1.05 \\
Estonia       & 0.30 & 1.84 & 0.96 \\
Cyprus        & 0.21 & 1.78 & 0.89 \\
Luxembourg    & 0.15 & 1.75 & 0.85 \\
Malta         & 0.12 & 1.74 & 0.83 \\

\toprule
\end{tabularx}
\label{Table3}
\end{table}

From Table~\ref{Table2}, we can observe the power indices of the EU’s medium-sized countries. These countries are relatively similar in both population and power, yet they possess significantly less influence over decisions compared to the larger member states mentioned earlier. As a result, their interests in the Union are primarily protected by the unanimity mechanism or their veto power in matters affecting national sovereignty.

Table~\ref{Table3} presents the countries with the smallest populations and lowest power indices. Compared to their population sizes, their power indices are proportionally higher. However, this still grants them  limited influence in decision-making. Among the small states, Luxembourg stands out—despite its small population, it plays a significant economic role in the EU and is a founding member. Thus, while its power index is low, in reality, it has numerous allies and greater influence.

We also assessed the power of various coalitions among small and medium-sized member states.  The Visegrád Four (V4) includes the Czech Republic, Hungary, Poland, and  Slovakia. Their cooperation is based on representing and aligning common political, economic, and diplomatic interests \citep{VisegradGroup2025}.


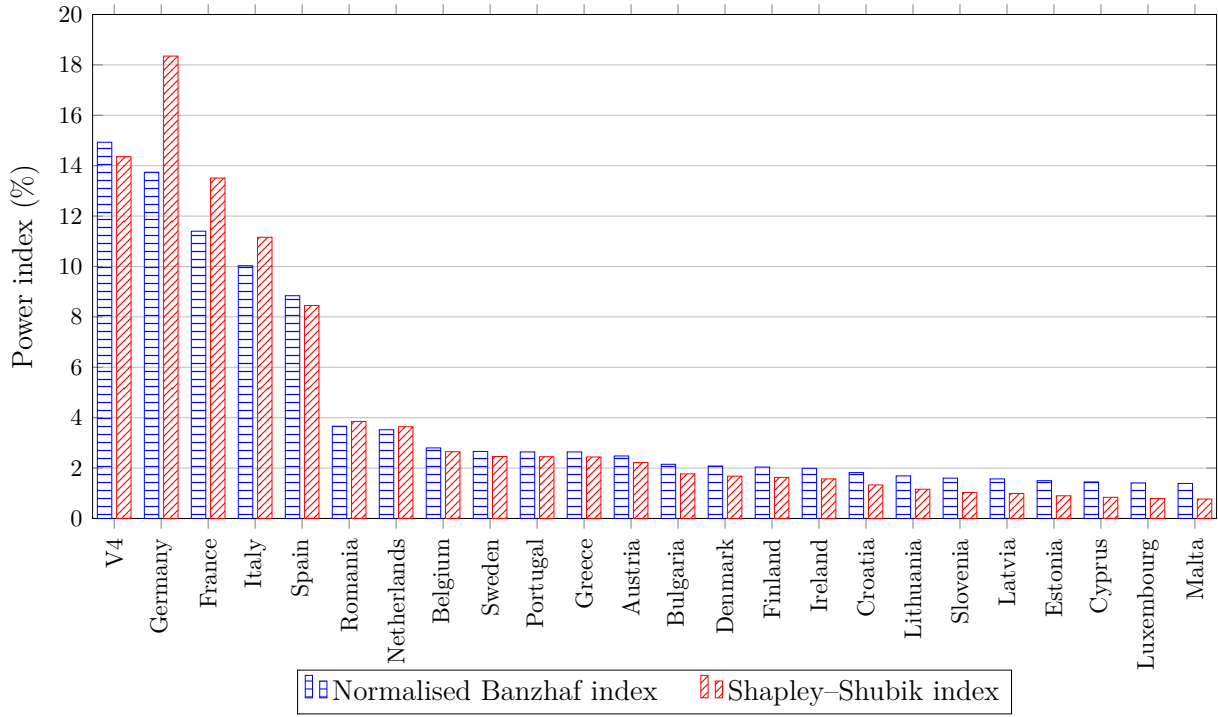
\begin{figure}[!t]
\centering
\begin{adjustbox}{width=\textwidth}
\begin{tikzpicture}
\begin{axis}[
    ybar,
    bar width=6pt,
    width=18cm,
    height=9cm,
    enlarge x limits=0.02,
    legend style={at={(0.5,-0.3)}, anchor=north, legend columns=-1},
    ylabel={Power index (\%)},
    symbolic x coords={V4,Germany,France,Italy,Spain,Romania,Netherlands,Belgium,Sweden,Portugal,Greece,Austria,Bulgaria,Denmark,Finland,Ireland,Croatia,Lithuania,Slovenia,Latvia,Estonia,Cyprus,Luxembourg,Malta},
    xtick=data,
    x tick label style={rotate=90, anchor=east},
    ymin=0,
    ymax=20,
    ymajorgrids = true,
    legend cell align={left}
]
\addplot+[style={blue,fill=blue!30, pattern=horizontal lines,  pattern color=blue}] coordinates {
(V4,14.93)
(Germany,13.74)
(France,11.4)
(Italy,10.03)
(Spain,8.84)
(Romania,3.66)
(Netherlands,3.52)
(Belgium,2.8)
(Sweden,2.66)
(Portugal,2.64)
(Greece,2.64)
(Austria,2.48)
(Bulgaria,2.15)
(Denmark,2.08)
(Finland,2.04)
(Ireland,1.99)
(Croatia,1.82)
(Lithuania,1.69)
(Slovenia,1.6)
(Latvia,1.57)
(Estonia,1.5)
(Cyprus,1.45)
(Luxembourg,1.41)
(Malta,1.39)
};
\addplot+[style={red,fill=red!30, pattern=north east lines, pattern color=red}] coordinates {
(V4,14.36)
(Germany,18.35)
(France,13.51)
(Italy,11.16)
(Spain,8.45)
(Romania,3.85)
(Netherlands,3.64)
(Belgium,2.65)
(Sweden,2.46)
(Portugal,2.45)
(Greece,2.44)
(Austria,2.22)
(Bulgaria,1.77)
(Denmark,1.68)
(Finland,1.63)
(Ireland,1.57)
(Croatia,1.33)
(Lithuania,1.16)
(Slovenia,1.03)
(Latvia,0.99)
(Estonia,0.9)
(Cyprus,0.84)
(Luxembourg,0.79)
(Malta,0.77)
};
\legend{Normalised Banzhaf index \quad\quad, Shapley--Shubik index}
\end{axis}
\end{tikzpicture}
\end{adjustbox}
\caption{Power indices with the cooperation of V4 countries}
\label{Figure4}
\end{figure}

Figure~\ref{Figure4} reports the results when the V4 countries vote as a coalition. Under the Shapley--Shubik index, Germany retains the highest power index, followed by the V4. However, under the Banzhaf index, the order is reversed—the V4, voting together in every decision, surpass Germany in power.

\citet{FreixasMarciniakPons2012} prove that the Banzhaf and Shapley--Shubik indices are ordinally equivalent in semicomplete games, which class contains (and is strictly larger than) the class of complete games and, therefore, the class of weighted games. Semicomplete games include binary voting systems such as the US federal system, which is not a complete game.  Thus, we have demonstrated an exceptional case, where the Banzhaf and Shapley--Shubik indices provide different rankings, which was caused by the double quota system (intersection of weighted voting games).


\begin{figure}[!t]
\centering
\begin{adjustbox}{width=\textwidth}
\begin{tikzpicture}
\begin{axis}[
    ybar,
    bar width=6pt,
    width=18cm,
    height=9cm,
    enlarge x limits=0.02,
    legend style={at={(0.5,-0.3)}, anchor=north, legend columns=-1},
    ylabel={Power index (\%)},
    symbolic x coords={2004,Germany,France,Italy,Spain,Romania,Netherlands,Belgium,Sweden,Portugal,Greece,Austria,Bulgaria,Denmark,Finland,Ireland,Croatia,Luxembourg},
    xtick=data,
    x tick label style={rotate=90, anchor=east},
    ymin=0,
    ymax=30,
    ymajorgrids = true,
    legend cell align={left}
]
\addplot+[style={blue,fill=blue!30, pattern=horizontal lines,  pattern color=blue}] coordinates {
(2004,25.2450951888076)
(Germany,17.7009315361624)
(France,14.0271203786618)
(Italy,12.154744321816)
(Spain,10.6917324711875)
(Romania,3.18362077988533)
(Netherlands,2.9487015479084)
(Belgium,1.99972088804121)
(Sweden,1.82643888029585)
(Portugal,1.81364624885157)
(Greece,1.80550548338702)
(Austria,1.59151964831893)
(Bulgaria,1.13214788281949)
(Denmark,1.05190319466896)
(Finland,0.993754869922197)
(Ireland,0.941421377650109)
(Croatia,0.722783676602277)
(Luxembourg,0.169211625013083)
};
\addplot+[style={red,fill=red!30, pattern=north east lines, pattern color=red}] coordinates {
(2004,26.7628449981391)
(Germany,16.5121969533734)
(France,12.0132563514916)
(Italy,10.109710550887)
(Spain,7.94533897475074)
(Romania,3.39038412567824)
(Netherlands,3.17697008873479)
(Belgium,2.37360678537149)
(Sweden,2.25109857462798)
(Portugal,2.24234915411386)
(Greece,2.23797444385679)
(Austria,2.09234393057922)
(Bulgaria,1.70883854707384)
(Denmark,1.65827636415871)
(Finland,1.62422544775485)
(Ireland,1.58185768479886)
(Croatia,1.37540564011152)
(Luxembourg,0.943321384497855)
};
\legend{Normalised Banzhaf index \quad\quad, Shapley--Shubik index}
\end{axis}
\end{tikzpicture}
\end{adjustbox}
\caption{Power indices with the cooperation of countries that joined in 2004}
\label{Figure5}
\end{figure}
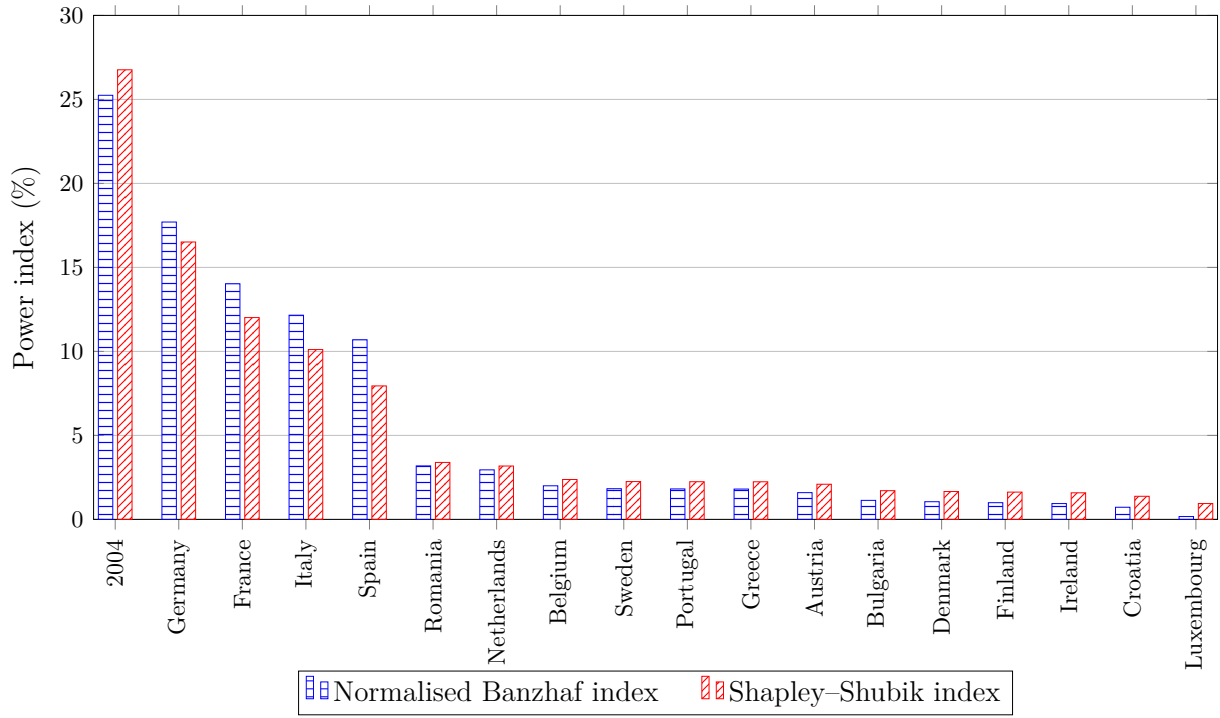
The analysis of countries that joined after 2004 is also interesting, as their accession marked the beginning of the current voting system. These countries include the Czech Republic, Cyprus, Estonia, Hungary, Latvia, Lithuania, Malta, Poland, Slovakia, and Slovenia. This enlargement united the Eastern Bloc with Western Europe and ended the post-World War II division of the continent. Figure~\ref{Figure5} shows the scenario when the 2004 entrants vote together and the rest vote individually.  

\begin{figure}[!t]
\centering
\begin{adjustbox}{width=\textwidth}
\begin{tikzpicture}
\begin{axis}[
    ybar,
    bar width=6pt,
    width=18cm,
    height=9cm,
    enlarge x limits=0.02,
    legend style={at={(0.5,-0.4)}, anchor=north, legend columns=-1},
    ylabel={Power index (\%)},
    symbolic x coords={Nordic,Germany,France,Italy,Spain,Poland,Romania,Netherlands,Belgium,Czech Republic,Portugal,Greece,Hungary,Austria,Bulgaria,Slovakia,Croatia,Slovenia,Cyprus,Luxembourg,Malta},
    xtick=data,
    x tick label style={rotate=90, anchor=east},
    ymin=0,
    ymax=20,
    ymajorgrids = true,
    legend cell align={left}
]
\addplot+[style={blue,fill=blue!30, pattern=horizontal lines,  pattern color=blue}] coordinates {
(Nordic,17.55)
(Germany,15.36)
(France,12.35)
(Italy,10.52)
(Spain,8.88)
(Poland,6.52)
(Romania,3.76)
(Netherlands,3.54)
(Belgium,2.55)
(Czech Republic,2.4)
(Portugal,2.33)
(Greece,2.32)
(Hungary,2.2)
(Austria,2.11)
(Bulgaria,1.67)
(Slovakia,1.5)
(Croatia,1.26)
(Slovenia,0.97)
(Cyprus,0.77)
(Luxembourg,0.73)
(Malta,0.71)
};
\addplot+[style={red,fill=red!30, pattern=north east lines, pattern color=red}] coordinates {
(Nordic,13.35)
(Germany,16.74)
(France,12.68)
(Italy,10.7)
(Spain,8.62)
(Poland,6.27)
(Romania,3.76)
(Netherlands,3.55)
(Belgium,2.65)
(Czech Republic,2.52)
(Portugal,2.47)
(Greece,2.46)
(Hungary,2.35)
(Austria,2.28)
(Bulgaria,1.9)
(Slovakia,1.77)
(Croatia,1.52)
(Slovenia,1.25)
(Cyprus,1.08)
(Luxembourg,1.05)
(Malta,1.03)
};
\legend{Normalised Banzhaf index \quad\quad, Shapley--Shubik index}
\end{axis}
\end{tikzpicture}
\end{adjustbox}
\caption{Power indices with the cooperation of Nordic countries}
\label{Figure6}
\end{figure}
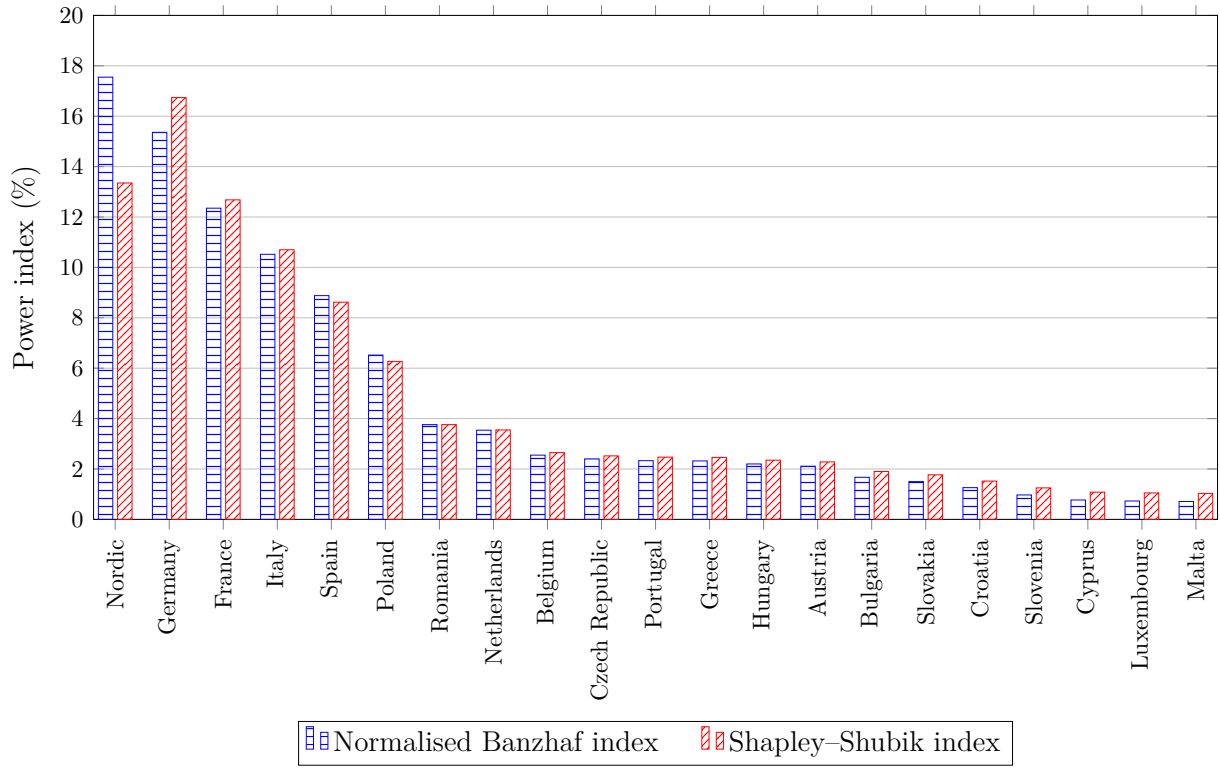
We also examine a regional coalition—the Nordic countries—which includes Denmark, Estonia, Finland, Latvia, Lithuania, and Sweden. Recently, tensions from the Russian invasion of Ukraine
have drawn these countries closer and pushed them toward deeper EU integration. Figure~\ref{Figure6} presents the power of this region. Similar to the V4, the results here are also contradictory: the Shapley--Shubik index gives Germany the most power, with the Nordics following France and Italy, while the Banzhaf index ranks the Nordic countries as the most powerful.

Overall, Germany and France hold the most power in the current European Union. However, if member states form coalitions, their influence can match or even exceed that of the most populous countries. 

\subsection{Enlargement with the Western Balkans}
In the following, we examine the changes in the balance of power within the Council if the EU expands to include the countries of the Western Balkans. Six countries in the Western Balkans are awaiting accession: Albania, Bosnia and Herzegovina, Kosovo, Montenegro, North Macedonia, and Serbia. We use demographic data from the OECD database, which is based on 2021 figures \citep{OECD2021}.

\begin{table}[t!]
\centering
\caption{Power indices of the acceding countries}
\rowcolors{0}{}{gray!25}
\begin{tabularx}{1.05\textwidth}{lCCC}\hiderowcolors
\toprule
Country & Population (\%) & Norm. Banzhaf index (\%) & Shapley--Shubik index (\%) \\
\bottomrule
\showrowcolors
Serbia & 1.49 & 2.36 & 1.77 \\
Bosnia and Herzegovina & 0.71 & 2.04 & 1.25 \\
Albania & 0.61 & 2.00 & 1.19 \\
North Macedonia & 0.45 & 1.93 & 1.08 \\
Kosovo & 0.38 & 1.91 & 1.04 \\
Montenegro & 0.13 & 1.80 & 0.87 \\

\toprule
\end{tabularx}
\label{Table4}
\end{table}

Table~\ref{Table4} shows the power indices and population data of the Western Balkan countries. We can see that these are small countries, the total population of the EU would not increase substantially—only by 3.9\%—while the member state quota would rise from 15 to 19. The small populations of the acceding countries, combined with the increased decision threshold, result in power indices that are disproportionately high relative to their population sizes.


\begin{figure}[!t]
\centering
\begin{adjustbox}{width=\textwidth}
\begin{tikzpicture}
\begin{axis}[
    ybar,
    bar width=6pt,
    width=18cm,
    height=9cm,
    enlarge x limits=0.02,
    legend style={at={(0.5,-0.4)}, anchor=north, legend columns=-1},
    ylabel={Change in power indices (percentage point)},
    symbolic x coords={Germany,France,Italy,Spain,Poland,Romania,Netherlands,Belgium,Czech Republic,Sweden,Portugal,Greece,Hungary,Austria,Bulgaria,Denmark,Finland,Slovakia,Ireland,Croatia,Lithuania,Slovenia,Latvia,Estonia,Cyprus,Luxembourg,Malta
},
    xtick=data,
    x tick label style={rotate=90, anchor=east},
    ymin=-3,
    ymax=0.15,
    ymajorgrids = true,
    legend cell align={left}
]
\addplot+[style={blue,fill=blue!30, pattern=horizontal lines,  pattern color=blue}] coordinates {
(Germany,-2.56)
(France,-2.05)
(Italy,-1.73)
(Spain,-1.44)
(Poland,-1.12)
(Romania,-0.52)
(Netherlands,-0.48)
(Belgium,-0.3)
(Czech Republic,-0.27)
(Sweden,-0.25)
(Portugal,-0.26)
(Greece,-0.26)
(Hungary,-0.22)
(Austria,-0.21)
(Bulgaria,-0.13)
(Denmark,-0.12)
(Finland,-0.1)
(Slovakia,-0.1)
(Ireland,-0.09)
(Croatia,-0.05)
(Lithuania,-0.02)
(Slovenia,0.01)
(Latvia,0.02)
(Estonia,0.03)
(Cyprus,0.05)
(Luxembourg,0.06)
(Malta,0.06)

};
\addplot+[style={red,fill=red!30, pattern=north east lines, pattern color=red}] coordinates {
(Germany,-1.65)
(France,-1.24)
(Italy,-1.04)
(Spain,-0.84)
(Poland,-0.57)
(Romania,-0.28)
(Netherlands,-0.25)
(Belgium,-0.15)
(Czech Republic,-0.13)
(Sweden,-0.14)
(Portugal,-0.14)
(Greece,-0.14)
(Hungary,-0.13)
(Austria,-0.13)
(Bulgaria,-0.08)
(Denmark,-0.08)
(Finland,-0.08)
(Slovakia,-0.08)
(Ireland,-0.07)
(Croatia,-0.04)
(Lithuania,-0.02)
(Slovenia,0)
(Latvia,0)
(Estonia,0.02)
(Cyprus,0.03)
(Luxembourg,0.03)
(Malta,0.03)

};
\legend{Normalised Banzhaf index \quad\quad, Shapley--Shubik index}
\end{axis}
\end{tikzpicture}
\end{adjustbox}
\caption{Changes in voting power following the EU33 enlargement}
\label{Figure7}
\end{figure}
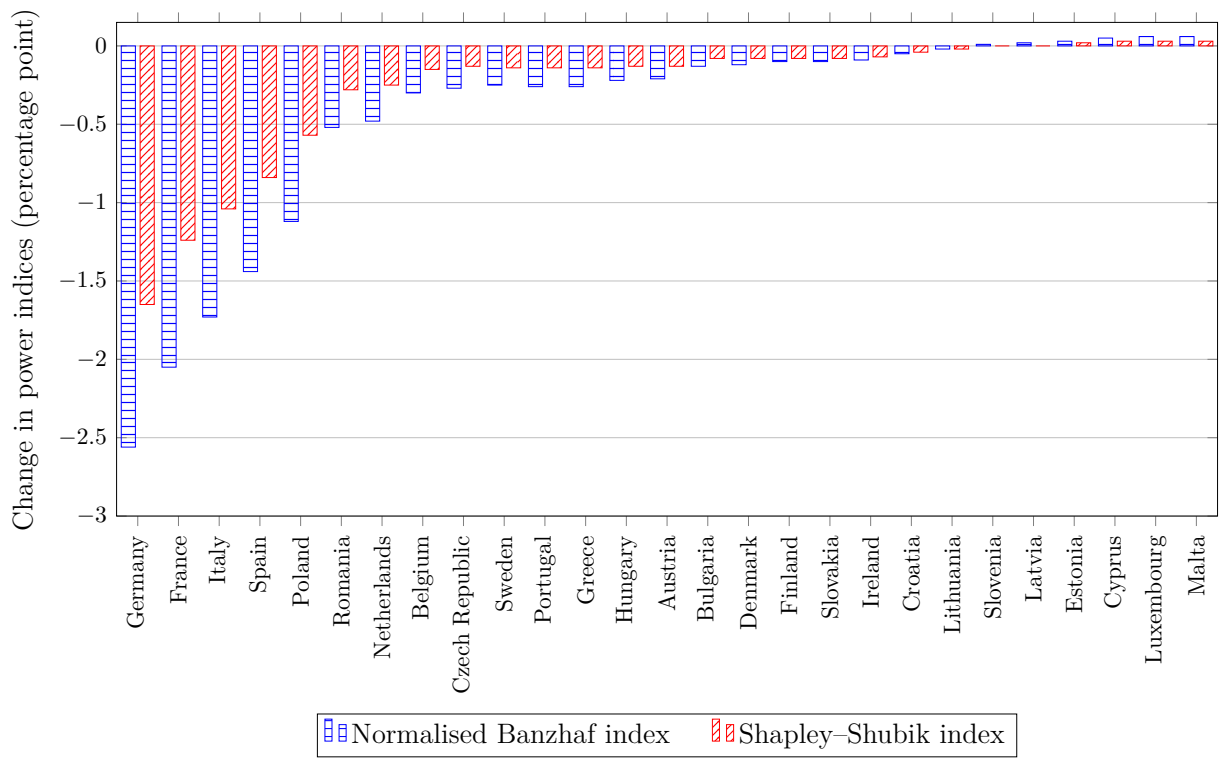
Figure~\ref{Figure7} illustrates the change between the power indices of the EU27 and the enlarged EU33 in percentage points. According to both power indices it is clear that countries with larger populations lose the most power. This can be explained by the previously mentioned case: the EU expanded with countries of smaller populations and the total population of the EU increased less than the member state quota. As a result, the phenomenon known as the "new member paradox" can be observed, where the enlargement causes a decrease in power for some member states, while others gain power. Thus, in the case of the EU33, countries with smaller populations see an increase in their power, while those with larger populations experience a decrease. Therefore, the smallest member states can be considered the winners of the enlargement based on the power indices, and they may have an interest in supporting further enlargement.

However, a difference can be observed between the results of the two indices. Latvia and Slovenia are in a special situation. The populations of most of the newly admitted countries are close to those of these two countries; therefore, while according to the Banzhaf index, Latvia gains 2 percentage points and Slovenia 1 percentage point, the Shapley--Shubik calculations show that their power indices remain unchanged. In other words, based on the Shapley--Shubik index, neither of them is worse off or better off.

\begin{table}[htbp]
  \centering
  \caption{Power indices of coalitions before and after enlargement with the Western Balkan countries}
  
  \rowcolors{2}{}{gray!25}
    \begin{tabularx}{\textwidth}{lccccCC}
    \toprule
    \multicolumn{1}{l}{\multirow{2}[4]{*}{Coalitions}} & \multicolumn{2}{c}{EU27} & \multicolumn{2}{c}{EU33} & \multicolumn{2}{c}{Rel. diff. (\%)} \\
\cmidrule{2-7}          & Banz. (\%) & S-S  (\%) & Banz. (\%) & S--S  (\%) & Banz. (\%) & S--S  (\%) \\
\bottomrule
   
    FRO and GER & 18.35 & 41.09 & 13.98 & 37.00 & $-$23.81 & $-$9.95  \\
   
    Weimar & 21.74 & 49.60 & 15.94 & 46.68 & $-$26.68 & $-$5.89  \\
    
    V4 & 14.93 & 14.36 & 12.49 & 13.13 & $-$16.34 & $-$8.57 \\
    
    2004 & 25.25 & 26.76 & 23.40 & 23.25 & $-$7.33 & $-$13.12 \\
    
    Founder & 37.99 & 58.64 & 24.76 & 53.35 & $-$34.82 & $-$9.02  \\
    
    Nordic & 14.95 & 10.72 & 13.85 & 10.05 & $-$7.36 & $-$6.25  \\
    \bottomrule
    \end{tabularx}%
  \label{Table5}%
\end{table}%

We also compute how the power of the coalitions assumed in the EU27 changes following the enlargement. Table~\ref{Table5} presents that after accession, all coalitions lose power. Coalitions that contain the founding member states experience the greatest relative losses according to the Banzhaf index; however, the relative difference is more minor if measured by Shapley--Shubik index. The Franco-German coalition and the Weimar Triangle partners are the next most affected, as they also include the largest countries. The Shapley--Shubik index shows that the greatest losers are the members of the 2004 enlargement.  Since the enlargement neither benefits any coalition nor has a neutral effect, this indicates a weakening of the stronger poles and a more polarized Union, where smaller states hold greater power than in the EU27.

As expected, the most populous countries lose the most power. In contrast, the states with the smallest populations would have a higher influence,  while Latvia and Slovenia remain neutral (according to the Shapley--Shubik index). Power loss can also be observed among the coalitions studied. Therefore, the balance of power shifts from the larger member states to the smaller ones due to the increased member state quota and the moderated population growth.

\subsection{Accession of Ukraine, Moldova, Georgia}

Ukraine's demographic situation and territorial boundaries are rather difficult to assess. This is due to the war, migration, and the absence of a recent comprehensive population census. However, population data remain crucial to compute the power indices. Therefore, we use Eurostat data, according to which Ukraine had nearly 41 million inhabitants in 2022. According to the same database, Moldova had slightly more than 2.5 million people in 2023. Georgia's population in 2023 was 3.7 million \citep{Eurostat2023}.

In the EU with 36 member states, the member state quota increases to 20, while the total population grows by 14.5\% compared to the EU27. In the case of the western Balkans, this population change is only 3.9\%.


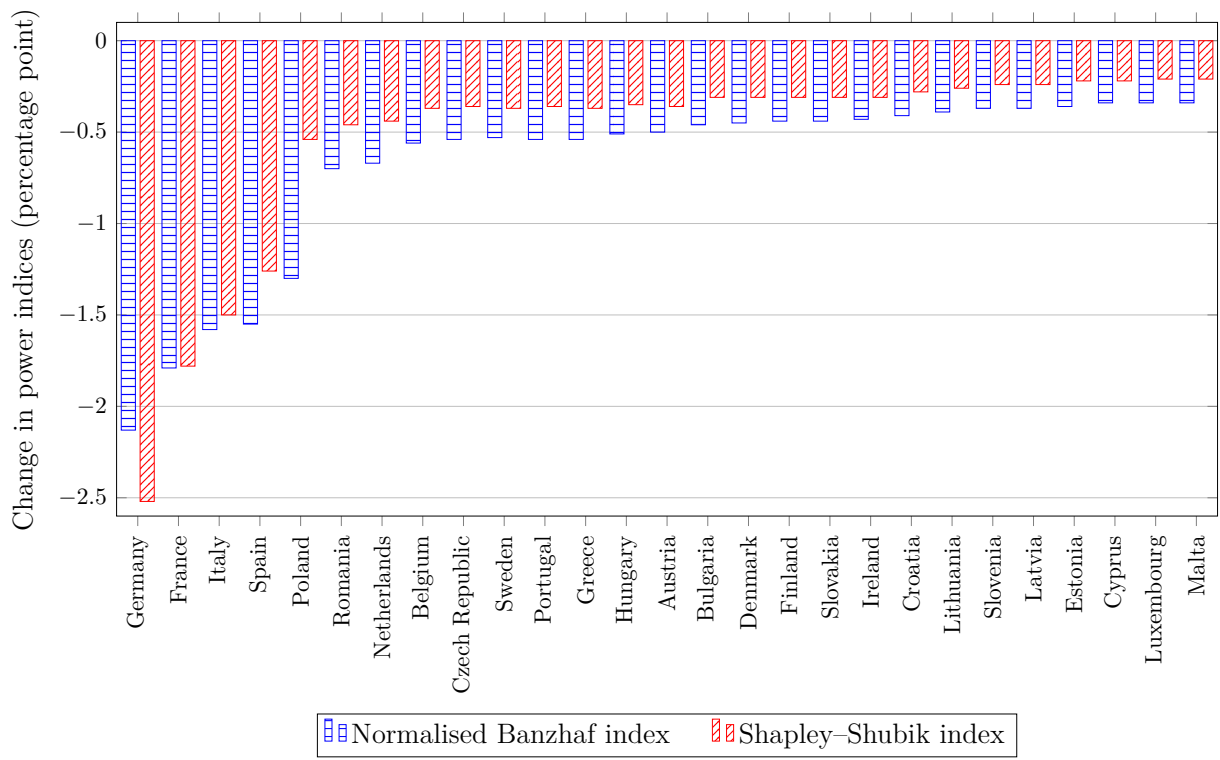
\begin{figure}[!t]
\centering
\begin{adjustbox}{width=\textwidth}
\begin{tikzpicture}
\begin{axis}[
    ybar,
    bar width=6pt,
    width=18cm,
    height=9cm,
    enlarge x limits=0.02,
    legend style={at={(0.5,-0.4)}, anchor=north, legend columns=-1},
    ylabel={Change in power indices (percentage point)},
    symbolic x coords={Germany,France,Italy,Spain,Poland,Romania,Netherlands,Belgium,Czech Republic,Sweden,Portugal,Greece,Hungary,Austria,Bulgaria,Denmark,Finland,Slovakia,Ireland,Croatia,Lithuania,Slovenia,Latvia,Estonia,Cyprus,Luxembourg,Malta
},
    xtick=data,
    x tick label style={rotate=90, anchor=east},
    ymin=-2.6,
    ymax=0.1,
    ymajorgrids = true,
    legend cell align={left}
]
\addplot+[style={blue,fill=blue!30, pattern=horizontal lines,  pattern color=blue}] coordinates {
(Germany,-2.13)
(France,-1.79)
(Italy,-1.58)
(Spain,-1.55)
(Poland,-1.3)
(Romania,-0.7)
(Netherlands,-0.67)
(Belgium,-0.56)
(Czech Republic,-0.54)
(Sweden,-0.53)
(Portugal,-0.54)
(Greece,-0.54)
(Hungary,-0.51)
(Austria,-0.5)
(Bulgaria,-0.46)
(Denmark,-0.45)
(Finland,-0.44)
(Slovakia,-0.44)
(Ireland,-0.43)
(Croatia,-0.41)
(Lithuania,-0.39)
(Slovenia,-0.37)
(Latvia,-0.37)
(Estonia,-0.36)
(Cyprus,-0.34)
(Luxembourg,-0.34)
(Malta,-0.34)

};
\addplot+[style={red,fill=red!30, pattern=north east lines, pattern color=red}] coordinates {
(Germany,-2.52)
(France,-1.78)
(Italy,-1.5)
(Spain,-1.26)
(Poland,-0.54)
(Romania,-0.46)
(Netherlands,-0.44)
(Belgium,-0.37)
(Czech Republic,-0.36)
(Sweden,-0.37)
(Portugal,-0.36)
(Greece,-0.37)
(Hungary,-0.35)
(Austria,-0.36)
(Bulgaria,-0.31)
(Denmark,-0.31)
(Finland,-0.31)
(Slovakia,-0.31)
(Ireland,-0.31)
(Croatia,-0.28)
(Lithuania,-0.26)
(Slovenia,-0.24)
(Latvia,-0.24)
(Estonia,-0.22)
(Cyprus,-0.22)
(Luxembourg,-0.21)
(Malta,-0.21)

};
\legend{Normalised Banzhaf index \quad\quad, Shapley--Shubik index}
\end{axis}
\end{tikzpicture}
\end{adjustbox}
\caption{Changes in voting power following the EU36 enlargement}
\label{Figure8}
\end{figure}
Figure~\ref{Figure8} illustrates the difference in power indices between the EU36 and the EU27 member states. According to both the Shapley-Shubik and Banzhaf indices, the countries with initially higher power indices lose the most from the enlargement. It can be observed that, based on the Shapley-Shubik index, Germany becomes the biggest loser of this accession, followed by France, Italy, and Spain. In the case of the Banzhaf index, Poland also ranks among the biggest losers. Furthermore, a notable difference arises compared to the accession of the Western Balkan countries, as in this scenario even the smaller member states do not gain, but ultimately lose in the EU36 configuration. Thus, this setup would not favor any of the EU27 member states based on the power indices.

\begin{table}[t!]
\centering
\caption{Data of the seven most populous countries in the 36-member Union}
\rowcolors{0}{}{gray!25}
\begin{tabularx}{1\textwidth}{lCcc}\hiderowcolors
\toprule
Country & Population (\%) & Norm. Banzhaf index (\%) & Shapley--Shubik index (\%) \\
\bottomrule
\showrowcolors
Germany        & 16.44 & 10.08 & 15.62 \\
France         & 13.26 & 8.29  & 11.82 \\
Italy          & 11.47 & 7.19  & 9.98  \\
Spain          & 9.36  & 6.14  & 8.01  \\
Ukraine        & 7.99  & 5.40  & 6.78  \\
Poland         & 7.16  & 4.91  & 6.10  \\

\toprule
\end{tabularx}
\label{Table6}
\end{table}

\begin{table}[t!]
\centering
\caption{Power indices of the EU 36 enlargement countries}
\rowcolors{0}{}{gray!25}
\begin{tabularx}{1\textwidth}{lcCC}\hiderowcolors
\toprule
Country & Population (\%) & Norm. Banzhaf index (\%) & Shapley--Shubik index (\%) \\
\bottomrule
\showrowcolors
Ukraine            & 7.99 & 5.40 & 6.78 \\
Serbia             & 1.35 & 2.04 & 1.54 \\
Georgia            & 0.73 & 1.72 & 1.08 \\
Bosnia and Herzegovina & 0.64 & 1.68 & 1.01 \\
Albania            & 0.56 & 1.63 & 0.95 \\
Moldova            & 0.49 & 1.60 & 0.90 \\
North Macedonia    & 0.41 & 1.55 & 0.84 \\
Kosovo             & 0.35 & 1.53 & 0.80 \\
Montenegro         & 0.12 & 1.41 & 0.63 \\

\toprule
\end{tabularx}
\label{Table7}
\end{table}
Table~\ref{Table6} presents the seven most populous countries in the 36-member Union and their resulting power indices. In this arrangement, Ukraine is ranked fifth, surpassing the power indices of 31 other countries. Consequently, the population weight and power of Germany and France decrease significantly. In the EU27 setup, the combined population of Germany and France accounts for 34\% of the EU27 population; however, in the EU36 scenario, this decreases to only 29.7\%, while with the simple accession of the Western Balkans, it amounts to 32.71\% of the EU33 population.

Table~\ref{Table7} illustrates the power indices of the nine acceding countries. The Western Balkan countries are disadvantaged in this case, as all of them lose power in their indices compared to Table~\ref{Table4}.

\begin{table}[!t]
  \centering
  \caption{Power indices of coalitions before and after the two enlargements}
  
  \rowcolors{2}{}{gray!25}
    \begin{tabularx}{\textwidth}{lcccccC}
    \toprule
    \multicolumn{1}{l}{\multirow{2}[4]{*}{Coalitions}} & \multicolumn{2}{c}{EU27} & \multicolumn{2}{c}{EU36} & \multicolumn{2}{c}{Diff. (\%)} \\
\cmidrule{2-7}          & Banz. (\%) & S--S  (\%) & Banz. (\%) & S--S  (\%) & Banz. (\%) & S--S  (\%) \\
\bottomrule
   
  FRO and GER & 18.35 & 41.09 & 14.59 & 33.28 & $-$20.49 & $-$19.01 \\
 Weimar & 21.74 & 49.6  & 16.69 & 47.42 & $-$23.23 & $-$4.40 \\
  V4 & 14.93 & 14.36 & 11.53 & 12.1  & $-$22.77 & $-$15.74 \\
  2004 & 25.25 & 26.76 & 20.65 & 19.26 & $-$18.22 & $-$28.03 \\
  Founder & 37.99 & 58.64 & 25.84 & 53.65 & $-$31.98 & $-$8.51 \\
   Nordic & 14.95 & 10.72 & 11.79 & 8.22  & $-$21.14 & $-$23.32 \\
    \bottomrule
    \end{tabularx}%
  \label{Table8}%
\end{table}%

\begin{figure}[!t]
\begin{centering}

\includegraphics[width=1\textwidth]{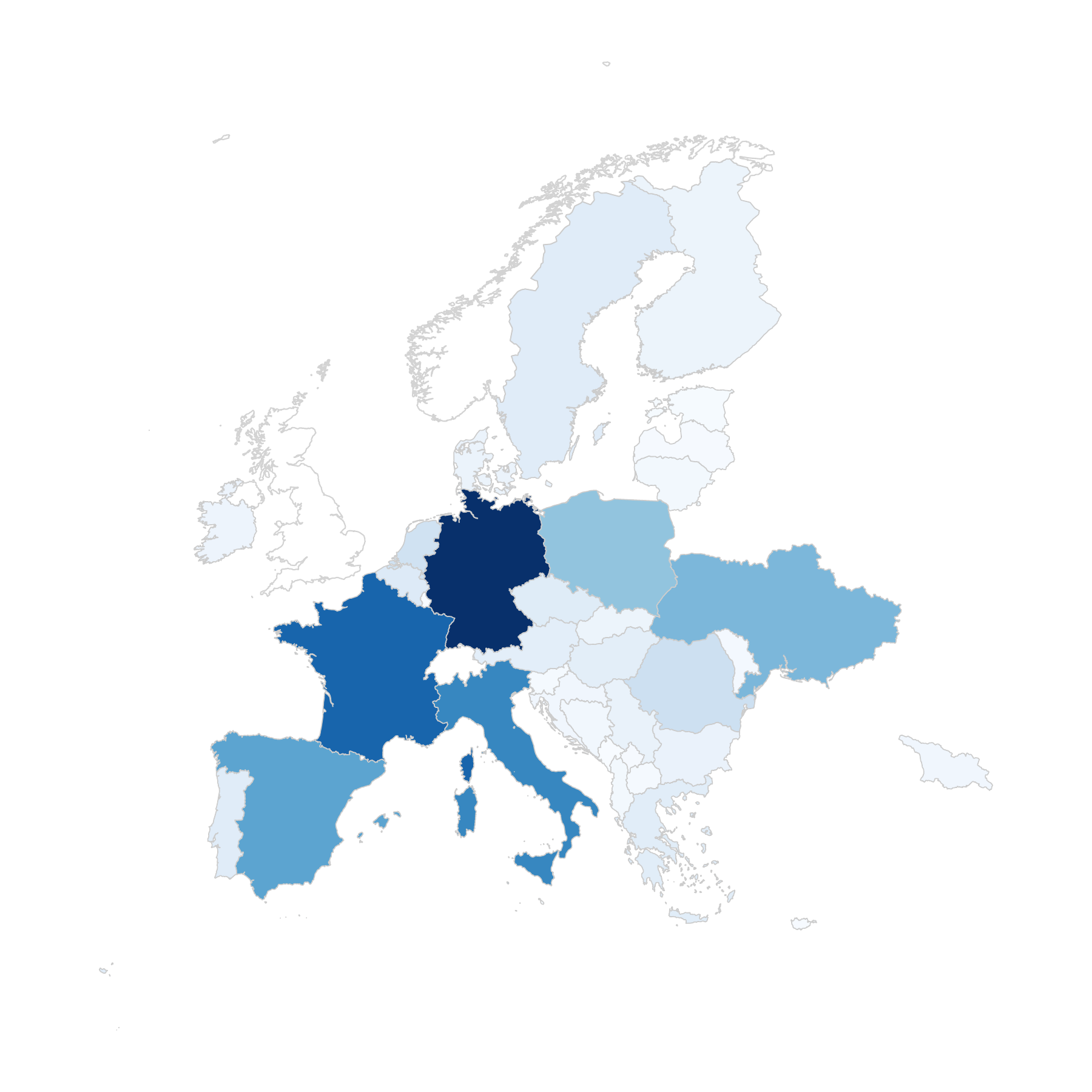}
\caption{Voting power in the EU36 measured by Banzhaf index}
\label{Figure9}
\end{centering}
\textit{Note: A darker colour indicates more power}

\end{figure}
Each of the examined coalitions loses power as illustrated in Table~\ref{Table8}. Similar to the accession of the Western Balkan countries, coalitions containing the most populous member states lose the most power.  However, the Franco-German, Weimar, and Founder coalitions, according to the Banzhaf index, are better off with EU36 than with EU33.

In the EU36, Ukraine would become the fifth most powerful member state, and the consequences of this are difficult to predict (see Figure~\ref{Figure9}). At the same time, all other member states would lose power, with the difficult losses affecting the countries that have been the most influential so far. This situation, where many lose power while Ukraine gains a strong influence, raises doubts about how many would actually support expanding the EU to 36 members under the current voting system. It also raises the question of whether the political support for Ukraine is just rhetoric or if it will truly lead to progress in the enlargement process

\section{Conclusions} \label{Sec5}

We observe two scenarios of enlargement. First, we examined the accession of the Western Balkan countries, resulting in an EU with 33 member states. Subsequently, we examined the accession of Georgia, Moldova and Ukraine the EU33, leading to a Union with 36 member states.

Our research found that the largest countries lose the most power in both the EU33 and EU36 scenarios, within the current institutional framework. Conversely, in the EU33 scenario, the phenomenon known as  new member paradox appears, meaning that while the power of the largest countries decreases, the power of the countries with the smallest populations increases. Thus, the enlargement of the Western Balkans would benefit the smallest countries while negatively impacting the largest countries.

In contrast, the EU36 scenario is disadvantageous for all countries. Ukraine would gain a strong position, becoming the country with the fifth-largest power.   

Regarding the examined coalitions, we also observed a loss of power in both enlargement scenarios. The most tremendous loss affects the founders, the Franco--German axis, and the Weimar Triangle, as these coalitions include the largest member states. Furthermore, we compared the value of the coalitions in the EU33 and EU36, finding that an EU36 enlargement would be more favourable for the large member states. This is because, in the EU33, the population increases by only 3.9\%, while the member state quota increases by four, requiring more member states to approve a proposal, with only a slight change in population data. This setup favours the small-population member states, whose power increases due to the member state quota requirement. In the EU36 scenario, the population increases by 14.5\%, meaning the population weight of the countries substantially decreases compared to the EU33. However, their member state quota changes less with an increase from 19 to 20.  The Franco--German, Weimar, and Founder coalitions, according to the Banzhaf index, lose less influence compared to the EU27, with the EU36 enlargement, rather than the EU33.

From a methodological perspective, it is essential to highlight that we found examples where the Banzhaf and the Shapley--Shubik indices do not produce the same ranking for either the V4 or the Nordic countries. For the V4, the Banzhaf index ranks the coalition as the strongest, while the Shapley--Shubik index ranks it as the second strongest after Germany. Similarly, for the Nordic countries, the Banzhaf index ranks them as having the highest power index, while the Shapley--Shubik index ranks them only fourth in the power hierarchy.

For further investigation, it would be interesting to see what Serbia's strength would mean if Bosnia and Herzegovina were evaluated according to entities and Republika Srpska and Serbia were included together. 

Our research takes into account only the institutional framework and is insensitive to such simple but important aspects as the individual qualities of leaders, personal good-bad relationships, historical alliances, and political alliances. Therefore, as a possible continuation of this research, we recommend investigating what anomalies and/or similarities can be found in practice in comparison.

\section*{Acknowledgements}
\addcontentsline{toc}{section}{Acknowledgements}
\noindent
We are grateful to \textit{László Csató}  for useful comments.

The research was supported by the National Research, Development and Innovation Office under Grants FK 145838 and PD 153835, and by the EK\"OP-24 University Research Scholarship Program of the Ministry for Culture and Innovation from the source of the National Research, Development and Innovation Fund. 

\bibliographystyle{apalike}
\bibliography{All_references}

\end{document}